\documentclass[preprint,12pt]{article}

\usepackage{amsmath,amssymb,amsfonts,latexsym,float,graphics,epsfig}

\usepackage{amsmath,amssymb,amsfonts,latexsym,float,graphics,epsfig}
\usepackage{verbatim}
\usepackage{sidecap}
\usepackage{graphicx}
\usepackage{subcaption}

\begin{document}

\renewcommand\theequation{\arabic{section}.\arabic{equation}}
\catcode`@=11 \@addtoreset{equation}{section}
\newtheorem{axiom}{Definition}[section]
\newtheorem{theorem}{Theorem}[section]
\newtheorem{axiom2}{Example}[section]
\newtheorem{lem}{Lemma}[section]
\newtheorem{prop}{Proposition}[section]
\newtheorem{cor}{Corollary}[section]
\newcommand{\lI}{\lambda_I}
\newcommand{\lR}{\lambda_R}
\newcommand{\be}{\begin{equation}}
\newcommand{\ee}{\end{equation}}
\newcommand{\bea}{\begin{eqnarray}}
\newcommand{\eea}{\end{eqnarray}}
\newcommand{\nn}{\nonumber}

\newcommand{\leqnomode}{\tagsleft@true}
\newcommand{\reqnomode}{\tagsleft@false}

\newcommand{\equal}{\!\!\!&=&\!\!\!}
\newcommand{\rd}{\partial}
\newcommand{\g}{\hat {\cal G}}
\newcommand{\bo}{\bigodot}
\newcommand{\res}{\mathop{\mbox{\rm res}}}
\newcommand{\diag}{\mathop{\mbox{\rm diag}}}
\newcommand{\Tr}{\mathop{\mbox{\rm Tr}}}
\newcommand{\const}{\mbox{\rm const.}\;}

\newcommand{\wbar}{{\bar{w}}}
\newcommand{\phibar}{{\bar{\phi}}}
\newcommand{\Psibar}{{\bar{\Psi}}}
\newcommand{\bLambda}{{\bf \Lambda}}
\newcommand{\bDelta}{{\bf \Delta}}
\newcommand{\p}{\partial}
\newcommand{\om}{{\Omega \cal G}}
\newcommand{\ID}{{\mathbb{D}}}
\newcommand{\pr}{{\prime}}
\newcommand{\prr}{{\prime\prime}}
\newcommand{\prrr}{{\prime\prime\prime}}

\title{Forced Coupled Duffing Oscillators with Nonlinear Damping: Resonance and Antiresonance}
\author{Ankan Pandey\footnote{E-mail: ankan0506@gmail.com}\\
SN Bose National Centre for Basic Sciences \\
JD Block, Sector III, Salt Lake \\ Kolkata 700098,  India \\}

\date{}
\maketitle

\begin{abstract}
In this work, we investigate resonance and antiresonance behaviour in forced coupled Duffing oscillators with nonlinear damping. Further, we will analyse the parameter dependence of the frequency response and stability. In the course of all the analysis, emphasis shall be on how different damping mechanisms contrast against each other. 
\end{abstract}

\paragraph{Keywords:}
Nonlinear damping, Multi-scale analysis, Antiresonance, Duffing oscillator

\normalsize
\section{Introduction}
Physical systems are not isolated in nature and does not behave independently. This often results in dissipation of energy making the system non-conservative in operation. The dissipation leads to the damping of motion of the units involved making the system to come stall. Such influence are present in chemical, biological and other system as well. \\
The phenomenological modelling of damping is done in terms of velocity raised to some power, usually written in the form $v|v|^{p-1}$ where $p\in \mathbb{R}$ . In most application, only the linear term, denoting viscous damping, is assumed. This linear assumption often proves to be too simplistic and fail to represent dissipation in the system. Such viscous damping is valid for low velocity oscillations in fluid mediums, \cite{beards1995engineering}. For different scenarios separate damping exponent is considered or a combination of them. Among integer order damping exponent, the most commonly employed exponents are, $p=0$, to model \textit{Coulomb damping} or dry friction, \cite{lim1998dynamics}. This type of damping occurs when two machine parts rub against each other $p=1$, for the linear case, and $p=2$ case representing \textit{Quadratic damping} which is encountered in flows with high Reynolds number, \cite{falzarano1992application}. Damping of fractional orders as well as combinations of different damping mechanisms have been utilised in physical modelling as well.\\
The effect of nonlinear damping terms on the dynamics has been studied for quite some time. In \cite{ravindra1994stability} Ravindra et. al. studied the stability of a nonlinearly damped hard Duffing oscillator. Furthermore in \cite{ravindra1994role}, they investigated the role of nonlinear damping  on the onset of period-doubling chaos in soft Duffing oscillator. In \cite{sanjuan1999effect}, Sanju\`an et al. discussed the role of nonlinear damping on some properties of universal escape oscillator. Hamiltonian description of the quadratic damping has been discussed in \cite{pandey2017chiellini}. \\
Duffing oscillator equation is one most important and extensively studied nonlinear system owing to the multitude of nonlinear phenomenon that it demonstrate given its simple form. Periodically forced Duffing oscillator shows jump phenomenon exhibiting a frequency hysteresis behaviour. The role of linear damping in the dynamics of Duffing oscillator has been studied widely and much literature exist about the subject, \cite{kovacic2011Duffing}. However, the effects of nonlinear damping is comparatively less studied, and hence invites for more research.\\
A single degree of freedom nonlinear oscillator under periodic forcing shows single resonance peak in its amplitude frequency response when the forcing frequency matches the oscillator's frequency. This phenomenon occurs for linear oscillator under periodic forcing as well. When such oscillators are coupled with another oscillator without any forcing, the frequency response of the forced oscillator shows dips between two resonance, termed antiresonance. Such dips appears when the forcing action of the coupled oscillator cancels the external forcing on the driven oscillator, causing destructive interference. Antiresonance is an important phenomenon in the field of nonlinear dynamics and have been interest of many research topics, \cite{wahl1999significance,sames2014antiresonance,
lysyansky2011desynchronizing,hanson2007role,d2000use}. In \cite{wahl1999significance}, the significance of antiresonance frequencies in experimental structural analysis is studied. Antiresonance phase shift was observed in strongly coupled cavity QED, \cite{sames2014antiresonance}. In \cite{jothimurugan2016multiple}, Jothimurugan et al. considered n-coupled Duffing oscillators with periodic forcing on one of them. They showed the presence of multiple resonance and antiresonance frequencies. \\
The resonance and antiresonance phenomenons in coupled Duffing oscillator have been well studied for systems with linear damping. However, the effect of nonlinear damping in such phenomenons have never been studied, in the best of author's knowledge. In this work, we will consider nonlinearly damped coupled Duffing oscillators with one of them being periodically forced and investigate for various effects of nonlinear damping on the resonance and antiresonance structures.

\setcounter{equation}{1}

\section{Periodically driven coupled Duffing oscillator}
Let us consider the system of interest in the form
\bea
\ddot{x} + \epsilon d \dot{x}|\dot{x}|^p + x - \epsilon b x^3 &=& \epsilon \alpha y + \epsilon F cos( \omega t) \\
\ddot{y} + \epsilon d \dot{y}|\dot{y}|^p + y - \epsilon b y^3 &=& \epsilon \alpha x .
\label{eq:CDOmain}
\eea
The equation represents linearly coupled Duffing oscillators with one of them periodically forced. In the equation, $\alpha$ is coupling coefficient, $F$ and $\omega$ are forcing amplitude and forcing frequency respectively, and rest of the parameters are usual system parameters. The exponent $p$ provides nonlinearity in the damping and can take integer values, for the purpose of this work. For further references, oscillator with dynamics represented by $x(t)$ will be referred as oscillator-x and that by $y(t)$ as oscillator-y.  \\

\subsection*{Theoretical analysis}
In this section we will analyse \eqref{eq:CDOmain} using multi-scale perturbation analysis and study frequency response of the solution. \\
Following the usual multi-scale perturbation procedure, the solution (zeroth order solution) is assumed to be of the form
\bea
x(t,t) &=& A(\tau)\, cos(\omega t + \phi_A(\tau)) \nn \\
y(t,t) &=& B(\tau)\,cos(\omega t + \phi_B(\tau)), \label{eq:appxSol}
\eea
where $\tau=\epsilon t$ is the slow time scale, $A$ and $\phi$ are amplitude and phase of the periodic solution, respectively.

The damping term in \eqref{eq:CDOmain} with above solution can be difficult to handle given the non smooth nature of the term. Therefore, the damping term is approximated in the form of Fourier series as
\be (-A\omega Sin(\omega t + \phi))|A\omega Sin(\omega t + \phi)|^p = -C_p\, \omega\, A |A|^p |\omega|^p \,sin(\omega t + \phi) + HHT, \ee
where $C_p = \frac{2 \Gamma(\frac{p+3}{2})}{\sqrt{\pi}\Gamma(\frac{p+4}{2})}$ and $\Gamma(x)$ is Gamma function. 'HHT' refers to higher harmonic terms which are going to be suppressed owing to rotating wave approximation.
The mod around $A$ and $\omega$ could be dropped as they are assumed positive . Using the above approximation along with solution \eqref{eq:appxSol} in the first order correction equation and setting the secular terms to zero gives the slow flow equation 
\begin{subequations}\label{eq:APE}
\bea 
A' &=& -\frac{C_p}{2}\,d\,\omega^p A^{p+1} + \frac{\alpha\,B}{2 \omega}\,sin(\phi_A - \phi_B) + \frac{F}{2\,\omega} \,sin\phi_A \label{eq:APEa} \\
A \phi_A' &=& -\frac{\delta\,A}{2\omega} + \frac{3\,b\, A^3}{8 \omega} + \frac{\alpha\,B}{2 \omega}\,cos(\phi_A - \phi_B) + \frac{F}{2\,\omega} \,cos\phi_A \label{eq:APEb} \\ 
B' &=& -\frac{C_p}{2}\,d\,\omega^p B^{p+1} - \frac{\alpha\,A}{2 \omega}\,sin(\phi_A - \phi_B) \label{eq:APEc} \\
B \phi_B' &=& -\frac{\delta\,B}{2\omega} + \frac{3\,b\, B^3}{8 \omega} + \frac{\alpha\,A}{2 \omega}\,cos(\phi_A - \phi_B)  \label{eq:APEd}. 
\eea
\end{subequations} 

The derivatives in \eqref{eq:APE} are with respect to slow time $\tau$. The fixed points of amplitude-phase equations \eqref{eq:APE} corresponds to the amplitude and phase of the periodic solution of \eqref{eq:CDOmain} and the stability of these points will determine the stability of the periodic solution. To calculate the fixed points, we set $A'=0,\,B'=0,\,\phi_A'=0,\,\phi_B'=0$, which, after some algebraic manipulations, gives the amplitude-frequency response equation, 
\scriptsize
\begin{subequations}\label{eq:AFresp}
\bea
\bigg( \frac{C_p d \omega^{p+1}B^{p+1}}{2} \bigg)^2 + \bigg( \frac{\delta B}{2} - \frac{3 B^3 b}{8} \bigg)^2 &=& \frac{\alpha^2 A^2}{4}, \label{eq:AFrespB}\\
\bigg( \frac{C_p d\, \omega^{p+1}}{2 A}( A^{p+2} + B^{p+2}) \bigg)^2 + \bigg(\frac{\delta}{2 A}(A^2 - B^2) - \frac{3 b}{8 A}(A^4 - B^4) \bigg)^2 &=& \frac{F^2}{4}, \label{eq:AFresp2}
\eea
\end{subequations}
\normalsize
and phase-frequency response equation,
\be \phi = \phi_A - \phi_B = arctan \bigg( -4\frac{C_p d \omega^{p+1} B^p}{ 4\delta - 3 b B^2} \bigg), \label{eq:PFresp} \ee
\be \phi_A = arctan \bigg( 4\frac{C_p d\,\omega^{p+1}( A^{p+2} + B^{p+2})}{ 4 \delta (A^2 - B^2) - 3\,b(A^4 - B^4) } \bigg). \label{eq:phiA} \ee
The frequency response relations shows the dependence of amplitude/phase on the forcing frequency. The roots of these algebraic equations are fixed points of amplitude-phase equation which could solved algebraically. However, owing to complexity of the equation it is not practically possible to solve it analytically. To the rescue are numerical methods which are employed to obtain the roots for different parameter arrangements.\\

\begin{figure}[t]
\begin{subfigure}{0.32\textwidth}
\includegraphics[width=5cm,height=5cm,keepaspectratio]{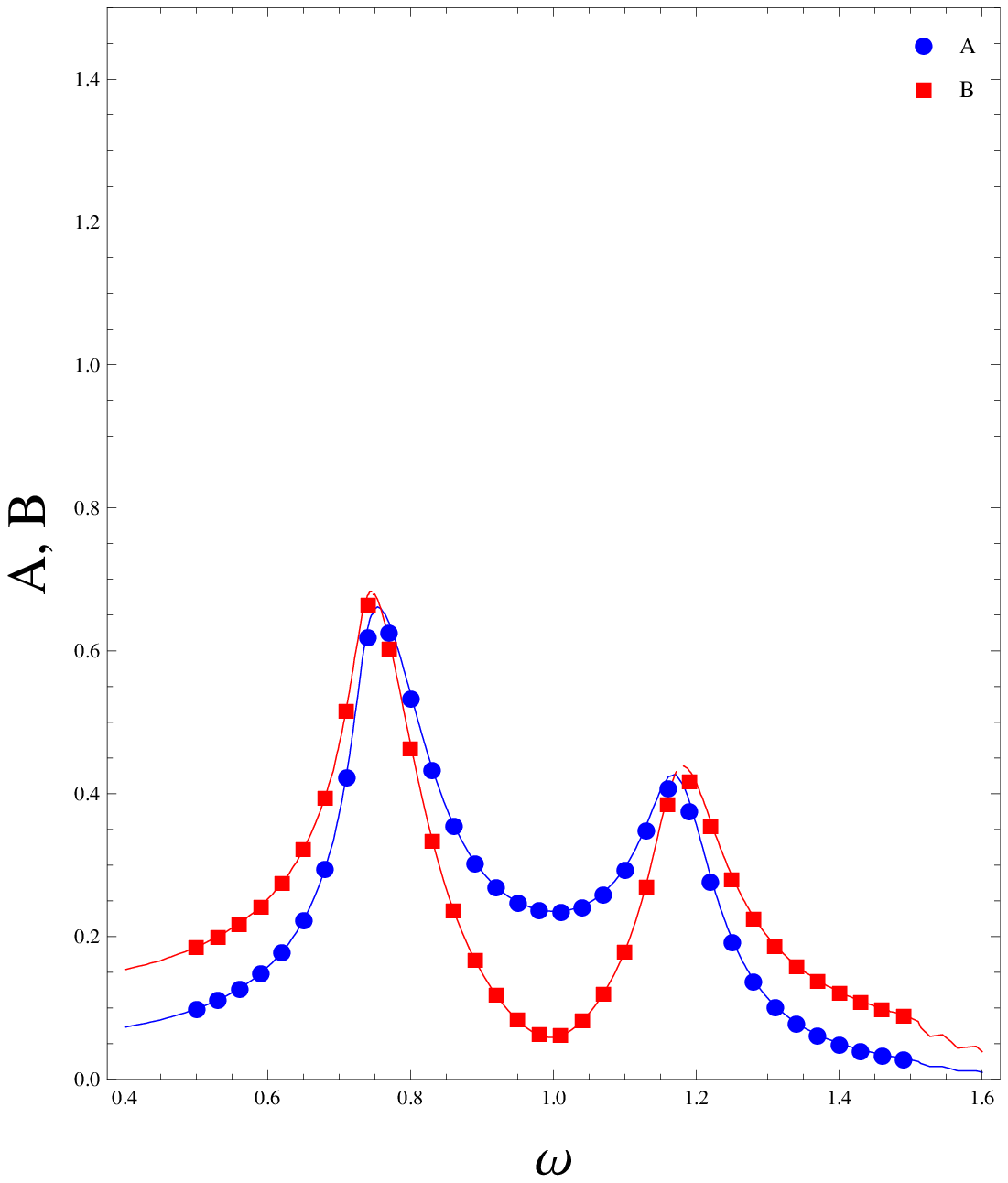}
\caption{$p=0$}
\end{subfigure}
\begin{subfigure}{0.32\textwidth}
\includegraphics[width=5cm,height=5cm,keepaspectratio]{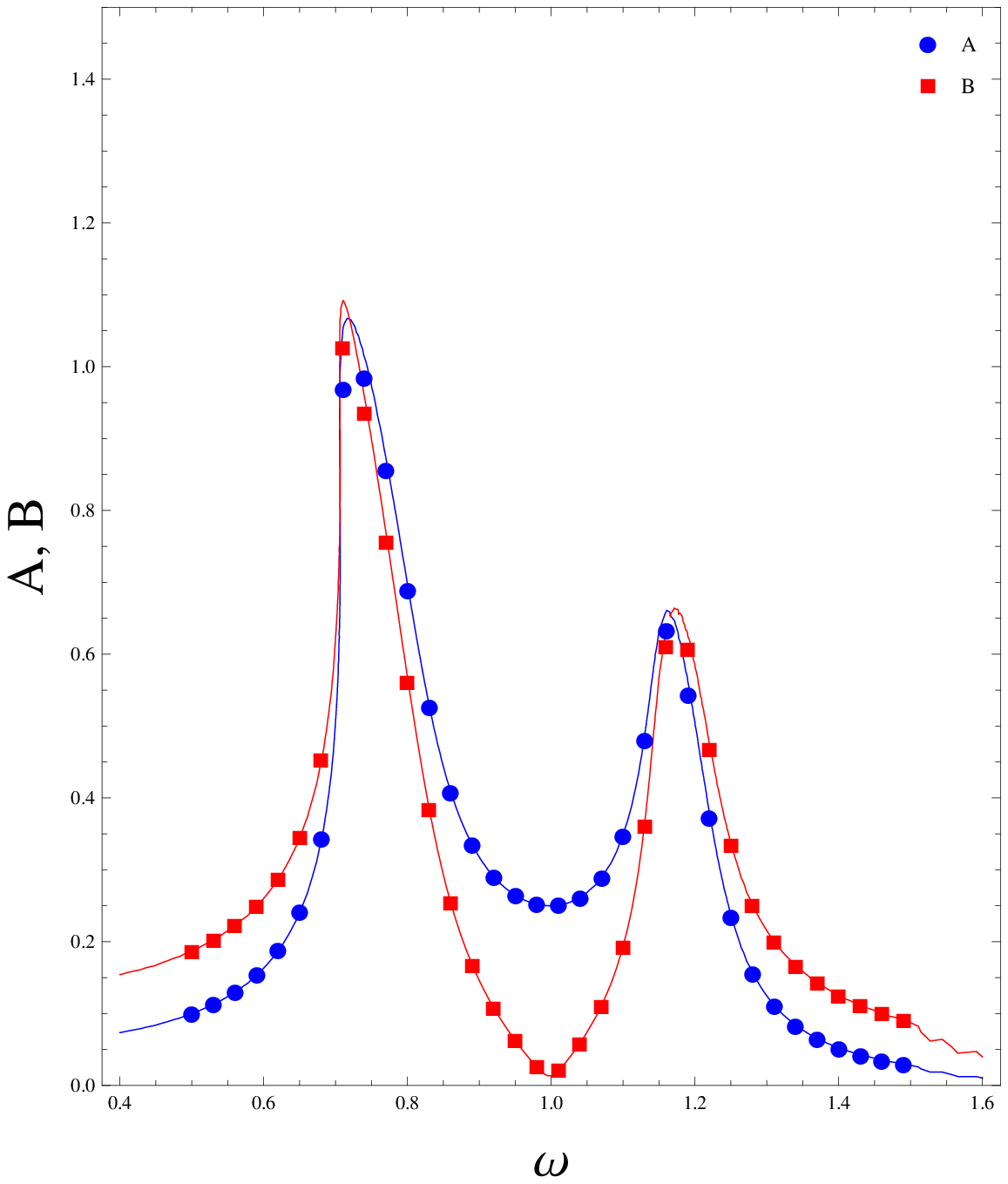}
\caption{$p=1$}
\end{subfigure}
\begin{subfigure}{0.32\textwidth}
\includegraphics[width=5cm,height=5cm,keepaspectratio]{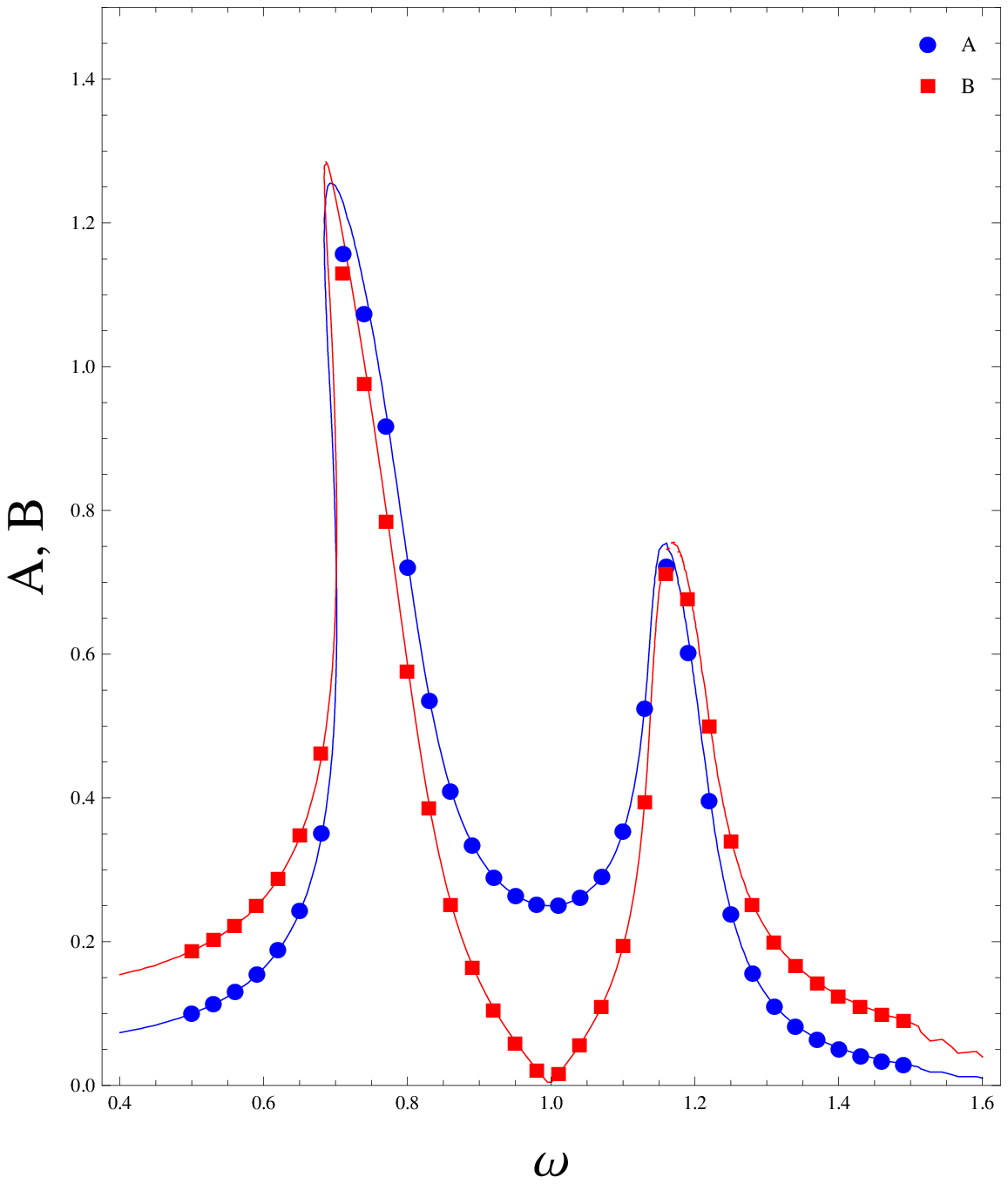}
\caption{$p=2$}
\end{subfigure}
\caption{Amplitude Frequency Response of amplitudes $A$ (of oscillator-x)(in blue) and $B$ (of oscillator-y)(in red). Three sub-plots are for different damping exponents: a) $p=0$; b) $p=1$; c) $p=2$. Continuous line represents theoretical predictions and solid circles and triangles are data points of numerically calculated amplitudes of $A$ and $B$, respectively. The parameter settings are: $d=1$, $b=1$, $\alpha=4$, $F=1$, $\epsilon=0.1$. }

\label{fig:AFR}
\end{figure}

\begin{figure}[!htb]
\begin{subfigure}{0.5\textwidth}
\includegraphics[width=6.5cm,height=6.5cm,keepaspectratio]{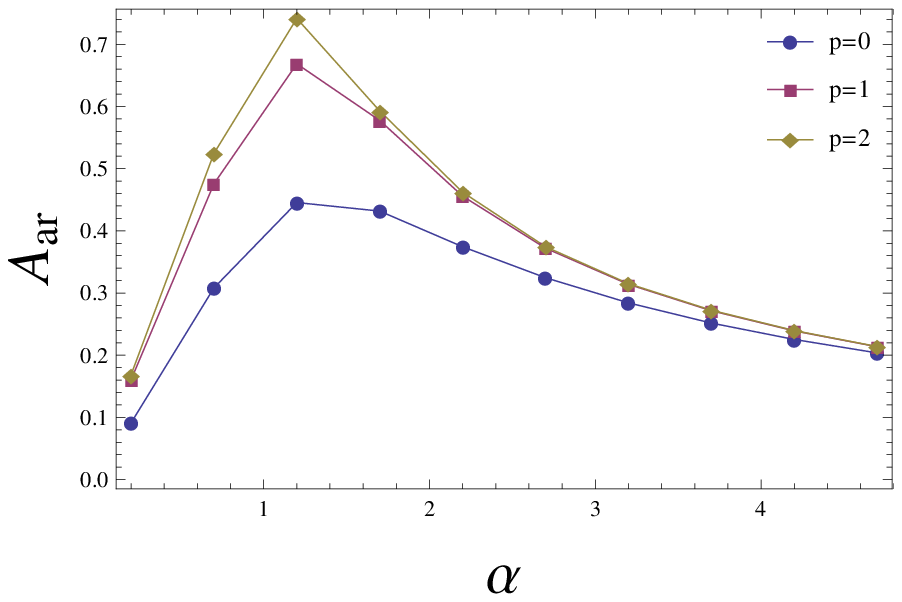}
\caption{$A_{ar}$ vs $\alpha$}
\end{subfigure}
\begin{subfigure}{0.5\textwidth}
\includegraphics[width=6.5cm,height=6.5cm,keepaspectratio]{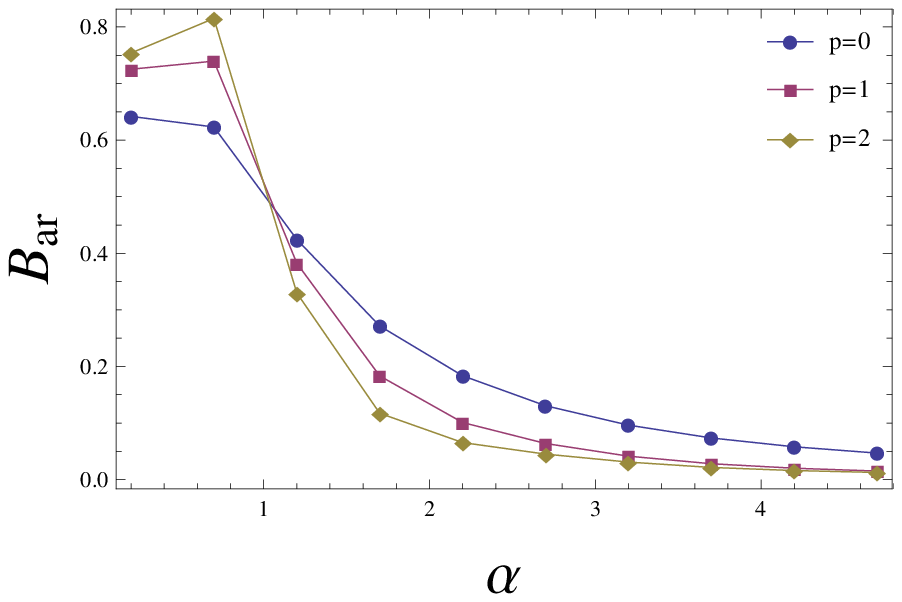}
\caption{$B_{ar}$ vs $\alpha$}
\end{subfigure}
\caption{The dependence of dips on coupling parameter $\alpha$ for damping exponents $p=0,1,2$. Antiresonance values for a) oscillator-x (amplitude $A$); b) oscillator-y (amplitude $B$). 
The data points are numerically computed points and continuous lines are trend lines. Shows separation between the three exponents for small $\alpha$ and convergence for large $\alpha$ values. The parameter settings are: $d=1$, $b=1$, $\alpha=4$, $F=1$, $\epsilon=0.1$. }

\label{fig:arVSalpha}
\end{figure}

 Figure\eqref{fig:AFR} shows the amplitude-frequency response curve together with numerically obtained data points by solving \eqref{eq:CDOmain} to show the authenticity of the theoretical results. The parameter values for the plot are: $d=1.0,\,b=1.0,\,F=1.0,\,\epsilon=0.1$, $\alpha=4.0$. This parameter setting is applicable on all the plots in this work, unless otherwise specified. The figure presents theoretical approximations calculated using perturbation analysis and numerical simulation calculations of the response function directly from \eqref{eq:CDOmain} which very verifies the theoretical predictions. The resonant peaks of individual oscillators will be refer to as $A_r^1$, $A_r^2$, and the corresponding frequency values, for resonance, as $\omega_r^{A1}$, and $\omega_r^{A2}$ and same goes for the oscillator-y with amplitudes replaced by $B$. The resonance in figure \eqref{fig:AFR} occurs at $\omega_r^{A1}=0.74$, $\omega_r^{A2}=1.19$ for $p=0$, $\omega_r^{A1}=0.71$, $\omega_r^{A2}=1.16$ for $p=1$, and $\omega_r^{A1}=0.71$, $\omega_r^{A2}=1.16$ for $p=2$. For oscillator-y the resonances occur at $\omega_r^{B1}=0.77$, $\omega_r^{B2}=1.16$ for $p=0$, $\omega_r^{B1}=0.74$, $\omega_r^{B2}=1.16$ for $p=1$, and $\omega_r^{B1}=0.71$, $\omega_r^{B2}=1.16$ for $p=2$. The resonance occurred at almost same frequencies regardless of the damping exponent. However, the calculations show that the amplitude value at resonance increases with the damping exponent as shown in the figure. This kind of increment in the maximum amplitude value is not expected and hence is non-trivial in nature.\\
The antiresonance dips occur at $\omega=1.0$ in all the cases for both the oscillators. Also, the dips for $p=1,2$ are shorter for small $\alpha$ but becomes identical with $p=0$ for larger values of coupling parameter $\alpha$. This dependence of antiresonance on $\alpha$ is shown in figure\eqref{fig:arVSalpha}. The parameter setting is same as that of previous computation.

\begin{figure}[t]
\begin{subfigure}{0.5\textwidth}
\includegraphics[width=5.5cm,height=5.5cm,keepaspectratio]{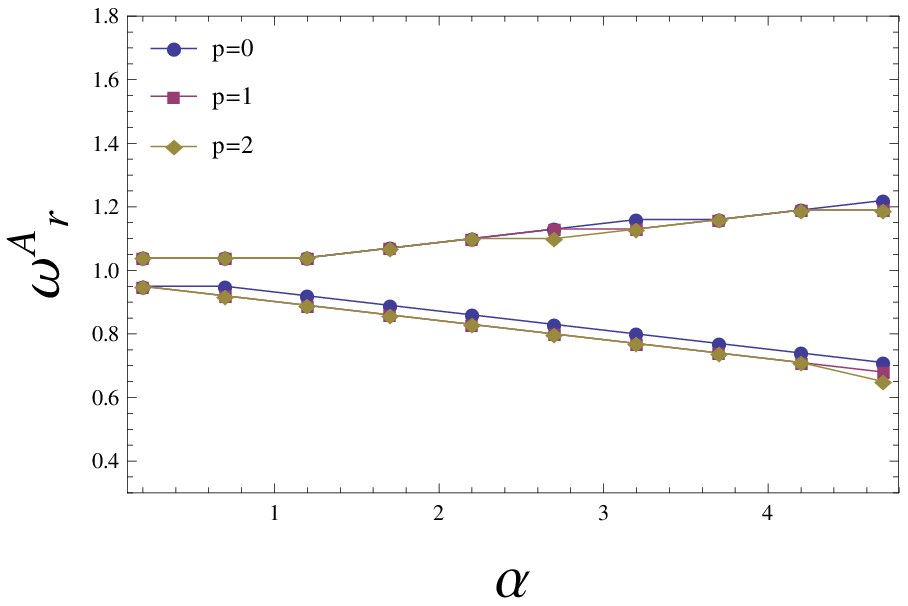}
\caption{$\omega_{r}^A$ vs $\alpha$}
\end{subfigure}
\begin{subfigure}{0.5\textwidth}
\includegraphics[width=5.5cm,height=5.5cm,keepaspectratio]{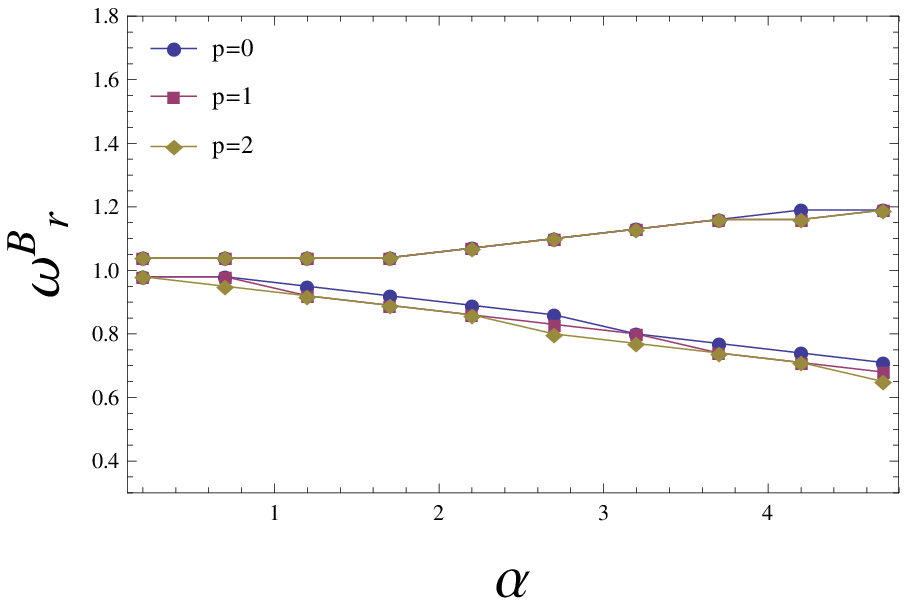}
\caption{$\omega_{r}^B$ vs $\alpha$}
\end{subfigure}
\caption{The dependence of resonant frequencies on coupling parameter $\alpha$ for damping exponents $p=0,1,2$. Resonance values for a) oscillator-x (amplitude $A$); b) oscillator-y (amplitude $B$). The data points are numerically computed points and continuous lines are trend lines. The two branches, below and above $\omega=1$, belongs to two resonances for each oscillator. The parameter settings are: $d=1$, $b=1$, $\alpha=4$, $F=1$, $\epsilon=0.1$. }

\label{fig:wrVSalpha}
\end{figure}   

\begin{figure}[!htb]
\begin{subfigure}{0.5\textwidth}
\includegraphics[width=5.5cm,height=5.5cm,keepaspectratio]{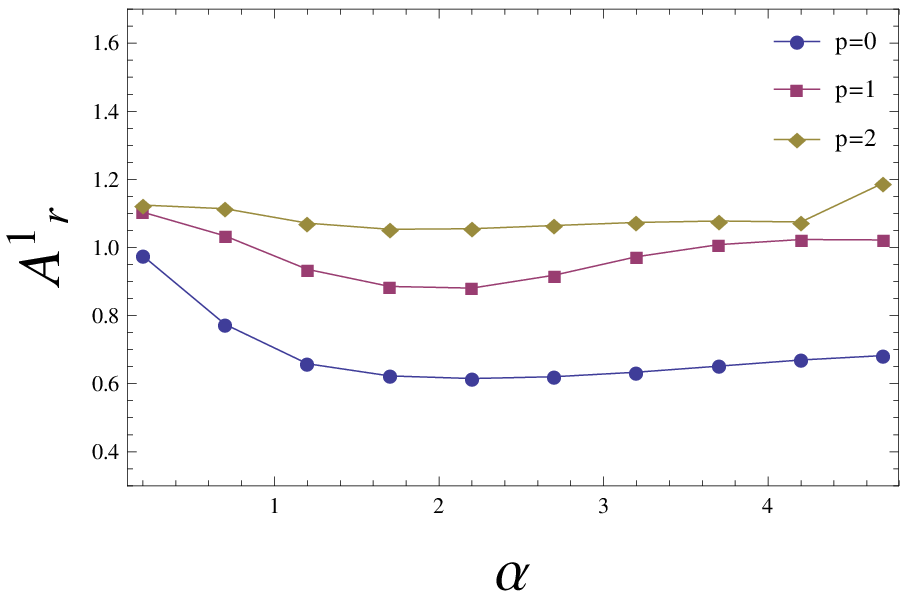}
\includegraphics[width=5.5cm,height=5.5cm,keepaspectratio]{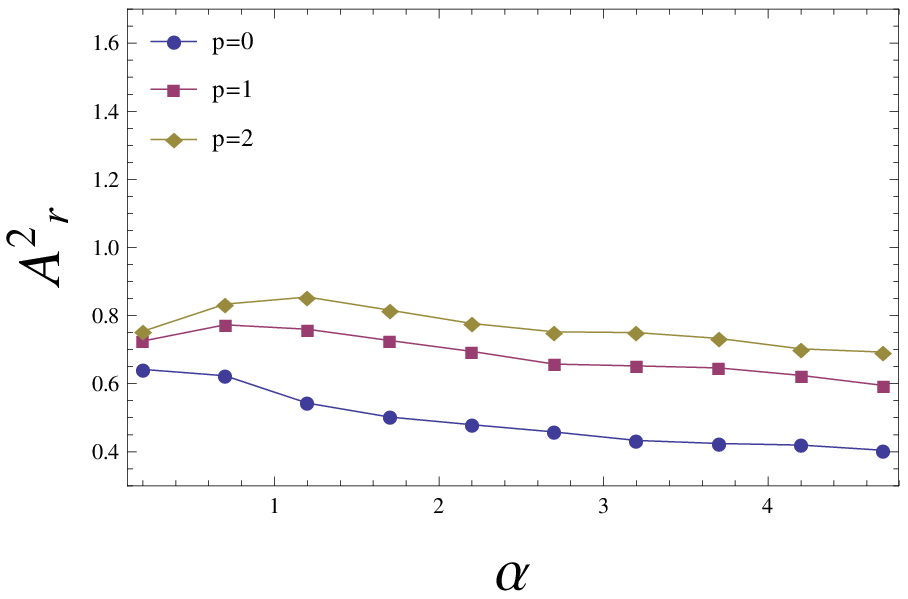}
\caption{$A_{r}$ vs $\alpha$}
\end{subfigure}
\begin{subfigure}{0.5\textwidth}
\includegraphics[width=5.5cm,height=5.5cm,keepaspectratio]{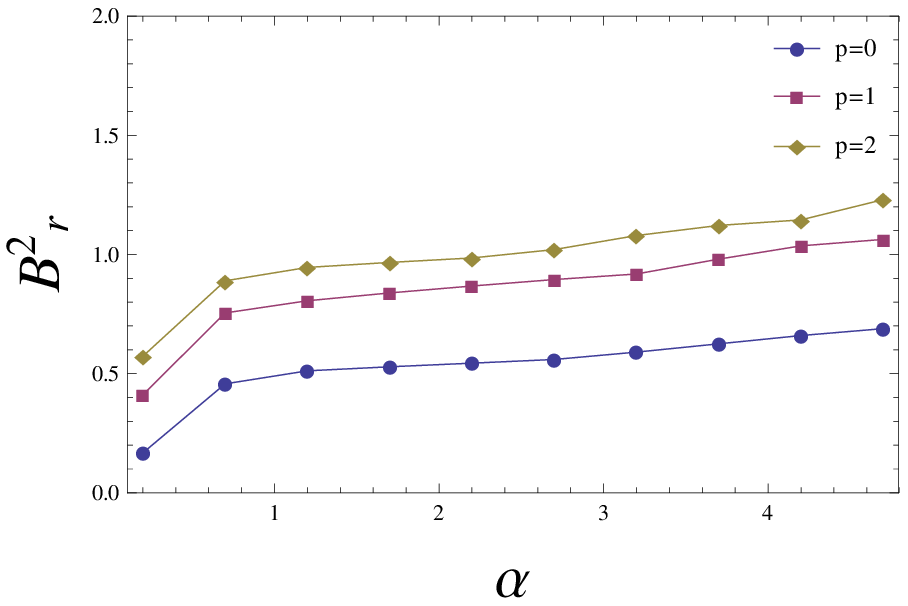}
\includegraphics[width=5.5cm,height=5.5cm,keepaspectratio]{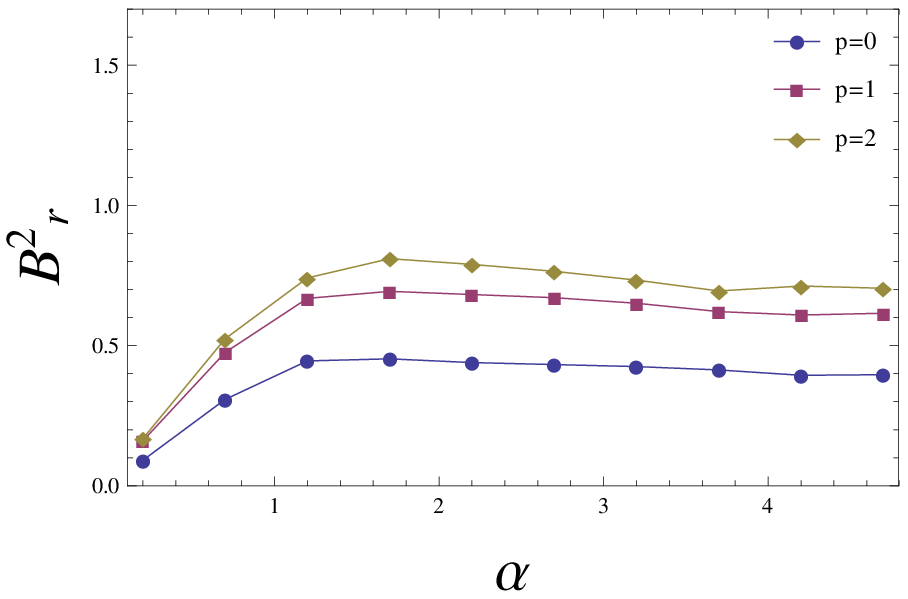}
\caption{$B_{r}$ vs $\alpha$}
\end{subfigure}
\caption{The dependence of resonant peaks on coupling parameter $\alpha$ for damping exponents $p=0,1,2$. Resonance values for a) oscillator-x (amplitude $A$); b) oscillator-y (amplitude $B$). The data points are numerically computed points and continuous lines are trend lines. The two branches, below and above $\omega=1$, belongs to two resonances for each oscillator. The parameter settings are: $d=1$, $b=1$, $\alpha=4$, $F=1$, $\epsilon=0.1$. }

\label{fig:rVSalpha}
\end{figure}

The resonance values for both the oscillators also depend on the coupling parameter. This dependence is shown in figure\eqref{fig:wrVSalpha}. The dependence is linear with respect to $\alpha$ and is same for all damping exponents. \\

The resonance peaks, as shown earlier, increase with the damping exponent as well as with coupling strength $\alpha$. This dependence is shown in figure\eqref{fig:rVSalpha}. The figure denotes a trend of saturation for large $\alpha$ for all the $p$ values. This trend is also present in the theoretical predictions.

\subsubsection*{Stability Analysis}
\begin{figure}[!htb]
\begin{center}
\includegraphics[width=15.5cm,height=10.5cm,keepaspectratio]{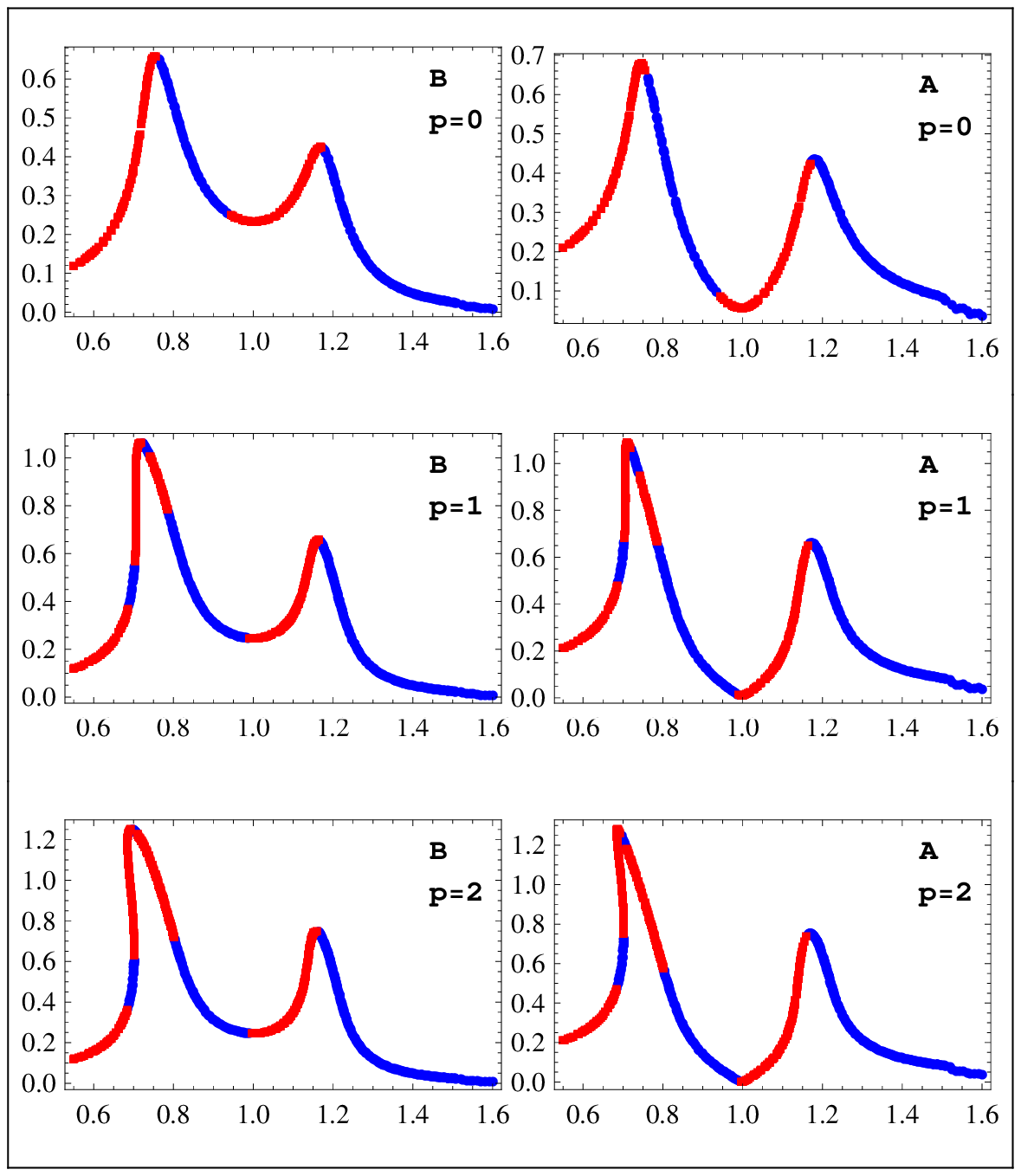}
\end{center}
\caption{Stability analysis of amplitude-phase equation \eqref{eq:APE}. The unstable values are shown in red and stable with blue. Each row corresponds to $p=0,\,p=1,\,p=2$ and shows stability of amplitudes $A$ and $B$. The parameter settings are: $d=1$, $b=1.5$, $\alpha=4$, $F=1$, $\epsilon=0.1$. }

\label{fig:stability}
\end{figure}

The stability of the periodic solutions corresponds to the stability of the fixed points of \eqref{eq:APE}. For the stability analysis, we did linear stability analysis of amplitude-phase dynamics. The stability plot is shown in figure\eqref{fig:stability}. From the analysis, it is evident that the unstable regions are increasing with damping exponent $p$. 

\subsubsection*{Phase analysis}

\begin{figure}[!htb]
\begin{subfigure}{0.5\textwidth}
\includegraphics[width=6.5cm,height=6.5cm,keepaspectratio]{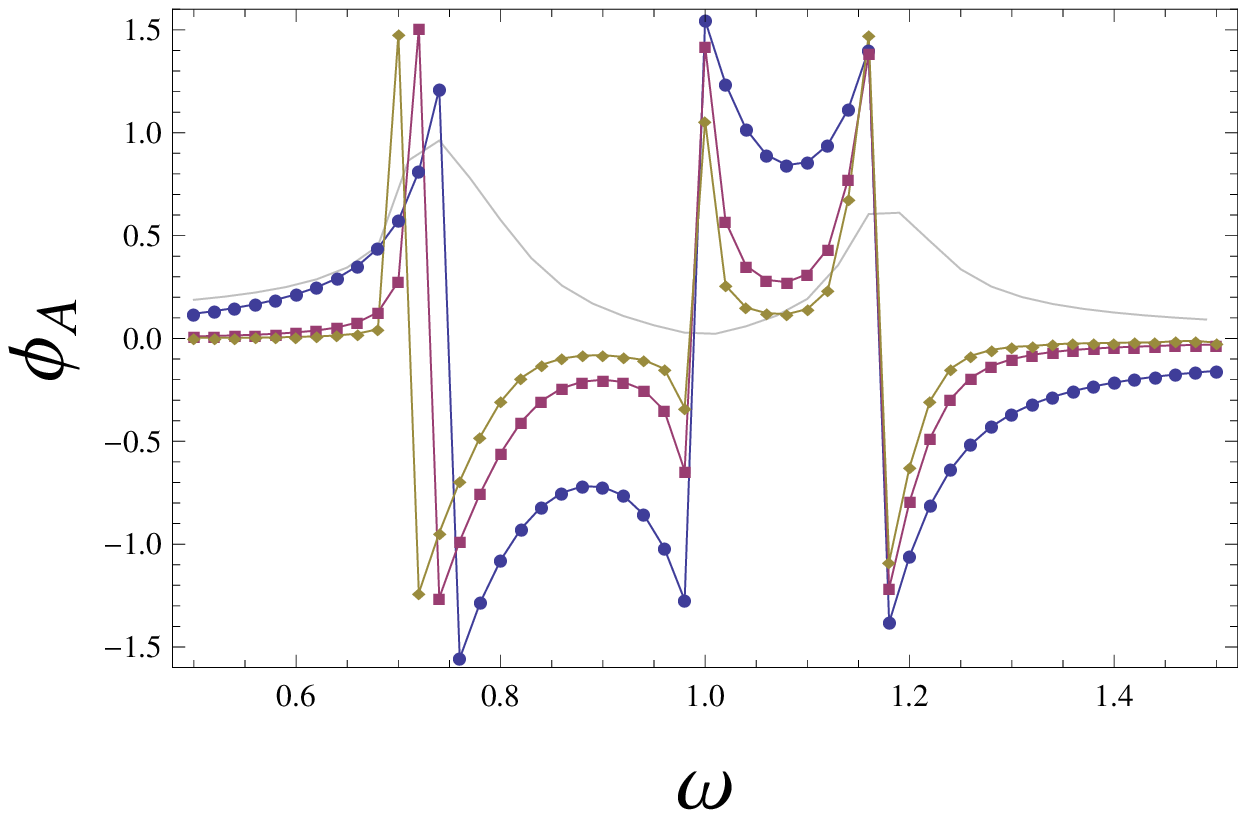}
\caption{ }
\label{fig:phaserespApa}
\end{subfigure}
\begin{subfigure}{0.5\textwidth}
\includegraphics[width=6.5cm,height=6.5cm,keepaspectratio]{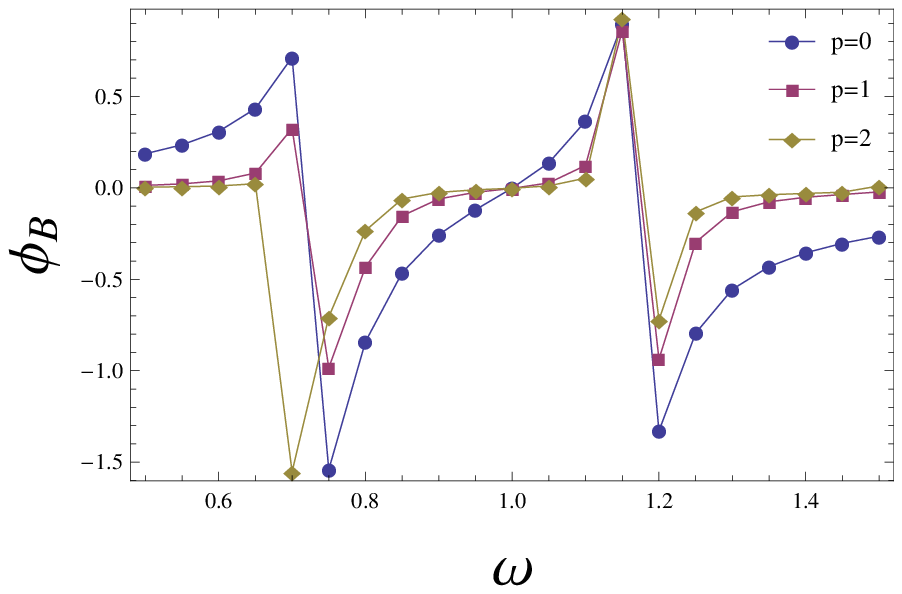}
\caption{ }
\label{fig:phaserespApb}
\end{subfigure}
\caption{Phase-frequency response of oscillator-x. a)The phase response of oscillator-x is given for all damping exponents. Amplitude response for $p=1$ is given gray to correlate the response events. This response is given for $p=2$. b) Phase response of oscillator-y for all the damping exponents. The parameter settings are: $d=1$, $b=1$, $\alpha=4$, $F=1$, $\epsilon=0.1$. }

\label{fig:phaserespAp}
\end{figure}

The phase dynamics is governed by \eqref{eq:APEc} and  \eqref{eq:APEd} and the fixed points are given by \eqref{eq:PFresp} and \eqref{eq:phiA}. The phase response is given in figure\eqref{fig:phaserespAp}. Oscillator-x shows an abrupt phase shift at antiresonance and resonance values for all the damping exponents. This is a characteristic of regular antiresonance in contrast to vibrational antiresonance where no such phase shift appears, \cite{sarkar2019vibrational}. For oscillator-y, there is no abrupt phase shift at antiresonance value for any $p$. The phase lag between the forcing and amplitude response decreases with damping exponent. This is evident from the phase response dependence on the $C_p$ term which decreases for with higher $p$ values.

\section{Discussion and Conclusion}

\begin{itemize}
\item[*] The frequency response of both the oscillators are obtained and vindicated by numerical simulation results. The intensity of each oscillator seems to be enhanced with the damping exponent. This observation is shown in figure \eqref{fig:AFR}. This is attributed to the fact that the coefficient of the first harmonic of the  Fourier series expansion of damping term decreases with the damping exponent. Hence with increasing damping exponents the contribution of the damping term decreases in the response function. This enhancement of amplitude denotes greater energy efficiency. 
\item[*] Dependence of the response at resonant and antiresonant values on coupling is obtained. Higher damping exponents show similar trend as that of linear damping with higher values. For oscillator-x the dependence of antiresonance amplitudes $A_{ar}$ take maxima at certain coupling and then amplitudes for all the exponents converge with increasing coupling. Oscillator-y amplitudes also converges for large coupling.\\
Antiresonance appears after certain coupling value for each oscillator following a pitchfork bifurcation in the amplitude phase dynamics. Afterwards, the two resonant peaks move away from each other with increasing coupling values in a linear fashion. This dependence is same for all the exponents.
 The resonant values saturates for large values of coupling for all the exponents.
\item[*] Stability of the periodic solutions of \eqref{eq:CDOmain} is obtained by analysing the stability of the fixed points of amplitude-phase equations \eqref{eq:APE}. The stability is determined by using linear stability analysis and the fixed points are, basically, frequency response values.  
\item[*] Phase lag between the forcing and the response decreases with increasing damping exponent. Phase response shows a phase shift in oscillator-x while no abrupt phase shift oscillator-y. This is a signature of antiresonance phenomenon. 
\end{itemize}

\section*{References}
\bibliographystyle{elsarticle-num}
\bibliography{coupmanuref}

\begin{thebibliography}{10}
\expandafter\ifx\csname url\endcsname\relax
  \def\url#1{\texttt{#1}}\fi
\expandafter\ifx\csname urlprefix\endcsname\relax\def\urlprefix{URL }\fi
\expandafter\ifx\csname href\endcsname\relax
  \def\href#1#2{#2} \def\path#1{#1}\fi

\bibitem{beards1995engineering}
C.~Beards, Engineering vibration analysis with application to control systems,
  Elsevier, 1995.

\bibitem{lim1998dynamics}
Y.~F. Lim, K.~Chen, Dynamics of dry friction: A numerical investigation,
  Physical Review E 58~(5) (1998) 5637.

\bibitem{falzarano1992application}
J.~M. Falzarano, S.~W. Shaw, A.~W. Troesch, Application of global methods for
  analyzing dynamical systems to ship rolling motion and capsizing,
  International journal of bifurcation and chaos 2~(01) (1992) 101--115.

\bibitem{ravindra1994stability}
B.~Ravindra, A.~Mallik, Stability analysis of a non-linearly damped duffing
  oscillator, Journal of Sound Vibration 171 (1994) 708--716.

\bibitem{ravindra1994role}
B.~Ravindra, A.~Mallik, Role of nonlinear dissipation in soft duffing
  oscillators, Physical review E 49~(6) (1994) 4950.

\bibitem{sanjuan1999effect}
M.~A. Sanjuan, The effect of nonlinear damping on the universal escape
  oscillator, International Journal of Bifurcation and Chaos 9~(04) (1999)
  735--744.

\bibitem{pandey2017chiellini}
A.~Pandey, A.~Ghose-Choudhury, P.~Guha, Chiellini integrability and
  quadratically damped oscillators, International Journal of Non-Linear
  Mechanics 92 (2017) 153--159.

\bibitem{kovacic2011Duffing}
I.~Kovacic, M.~J. Brennan, The Duffing equation: nonlinear oscillators and
  their behaviour, John Wiley \& Sons, 2011.

\bibitem{wahl1999significance}
F.~Wahl, G.~Schmidt, L.~Forrai, On the significance of antiresonance
  frequencies in experimental structural analysis, Journal of Sound and
  Vibration 219~(3) (1999) 379--394.

\bibitem{sames2014antiresonance}
C.~Sames, H.~Chibani, C.~Hamsen, P.~A. Altin, T.~Wilk, G.~Rempe, Antiresonance
  phase shift in strongly coupled cavity qed, Physical review letters 112~(4)
  (2014) 043601.

\bibitem{lysyansky2011desynchronizing}
B.~Lysyansky, O.~V. Popovych, P.~A. Tass, Desynchronizing anti-resonance effect
  of m: n on--off coordinated reset stimulation, Journal of Neural Engineering
  8~(3) (2011) 036019.

\bibitem{hanson2007role}
D.~Hanson, T.~Waters, D.~Thompson, R.~Randall, R.~Ford, The role of
  anti-resonance frequencies from operational modal analysis in finite element
  model updating, Mechanical Systems and Signal Processing 21~(1) (2007)
  74--97.

\bibitem{d2000use}
W.~D'AMBROGIO, A.~Fregolent, The use of antiresonances for robust model
  updating, Journal of Sound and Vibration 236~(2) (2000) 227--243.

\bibitem{jothimurugan2016multiple}
R.~Jothimurugan, K.~Thamilmaran, S.~Rajasekar, M.~Sanju{\'a}n, Multiple
  resonance and anti-resonance in coupled duffing oscillators, Nonlinear
  Dynamics 83~(4) (2016) 1803--1814.

\bibitem{sarkar2019vibrational}
P.~Sarkar, D.~S. Ray, Vibrational antiresonance in nonlinear coupled systems,
  Physical Review E 99~(5) (2019) 052221.

\end{thebibliography}
 
\end{document}